\documentclass[a4paper,11pt]{article}
\usepackage{jheppub}
\usepackage{amssymb,amsmath,mathrsfs,enumerate}
\usepackage{graphicx,rotate,multicol}
\usepackage{float}

\usepackage{subfig}

\usepackage{slashed}
\usepackage{mathtools}
\usepackage{multirow}

\usepackage{orcidlink}

\allowdisplaybreaks

\title{\boldmath Enlightening dark moments of neutrino with superradiance}

\author[a,b]{Indra Kumar Banerjee,$^{\orcidlink{https://orcid.org/0000-0003-3900-735X},}$}
\emailAdd{indrakumarb@imsc.res.in}

\author[c]{Ujjal Kumar Dey,$^{\orcidlink{https://orcid.org/0000-0002-9620-7561},}$}
\emailAdd{ujjal@iiserbpr.ac.in}

\author[c]{Anna John$^{\orcidlink{https://orcid.org/0009-0003-1871-3519},}$}
\emailAdd{anna23@iiserbpr.ac.in}

\affiliation[a]{Institute of Mathematical Sciences, \\ HBNI, CIT Campus, Chennai 600113, India}

\affiliation[b]{Homi Bhabha National Institute,\\Training School Complex, Anushakti Nagar, Mumbai 400085, India}

\affiliation[c]{Department of Physical Sciences, Indian Institute of Science Education and Research Berhampur,\\Berhampur, Odisha 760003, India}

\abstract{Neutrinos can acquire electromagnetic moments either within the Standard Model through higher order radiative corrections or within the domain of new physics. In this study we focus on probing these beyond the standard model neutrino moments through quenched superradiance of black holes where fermionic pairs can be produced from the superradiant bosonic cloud. We consider the production of dark photons from black hole superradiance and quenching occurs through the production of neutrino-antineutrino pairs from the dark photons. The efficiency of the pair production depends on the effective coupling between the dark photons and neutrinos, i.e., the dark electromagnetic moments. We also discuss bounds on primordial black hole abundance from neutrino background arising from this quenched superradiance mechanism.}

\begin{document}
\maketitle
\flushbottom

\section{Introduction}
\label{sec:intro}
The Standard Model (SM) of particle physics describes neutrinos as the massless, electrically neutral leptons. They are considered to be one of the best messengers for astrophysical and cosmological studies due to their ability to travel extremely long distances without significant interactions. However, SM is inadequate to answer various questions regarding neutrinos, such as the origin of neutrino mass mandated by the oscillation experiments. The concept of neutrino mass led to extensive research in different aspects such as the nature of the neutrinos as Dirac or Majorana particles, neutrino decay, and various neutrino interactions. Considering the fact that neutrinos are electrically neutral, they were expected to not take part in electromagnetic interactions. However, even though these interactions are forbidden at the tree level, it was found that neutrinos could obtain electromagnetic moments through loop corrections not only within the purview of SM, but also from beyond the Standard Model (BSM) and extensive research has been carried out in this area \cite{Nieves:1981zt,Dolgov:1981hv,Kayser:1982br,Bilenky:1987ty,RAFFELT19901,Pulido:1991fb, Salati:1993tf, Raffelt:1999gv, Raffelt:1999tx,Raffelt:2000kp, Dolgov:2002wy, Nowakowski:2004cv, Wong:2005pa, Giunti:2008ve, Studenikin:2008bd, Broggini:2012df, Giunti:2014ixa, Akhmedov:2014kxa}. In the former, the SM fields take part in the loop level interaction between photons and neutrinos whereas in the latter, depending on the model, along with the SM fields, various BSM fields such as BSM fermions and dark photons facilitate the interaction.
%

%
The dark photon is a hypothetical gauge boson corresponding to BSM $U(1)$ gauge symmetry \cite{Georgi:1983sy,Holdom:1985ag,Raggi:2015yfk,Deliyergiyev:2015oxa,Curciarello:2016jbz,Fabbrichesi:2020wbt, Caputo:2026pdw}. In addition to that, dark photons can have very weak interactions with the electrically charged particles though kinetic mixing ($\epsilon$) with the SM photon. Various models of dark photons can be found in literature where dark photons can be either massive or massless \cite{Pospelov:2007mp,Redondo:2008ec,Nelson:2011sf,Vogel:2013raa,Fabbrichesi:2020wbt, Ennadifi:2022vle,Berlin:2022hmt}. Furthermore, dark photons can themselves play the role of particle dark matter or they can mediate some interactions between other dark sector particles. 
In this study we consider a scenario where the neutrinos have loop level interactions with the dark photons and obtain their dark electromagnetic moments. However we remain agnostic regarding the other participants of the relevant interactions and therefore perform a model independent study.
Another interesting phenomenon driven by \textit{new physics} relevant to astroparticle physics is black hole superradiance\footnote{Superradiance can in principle occur around various compact objects \cite{Cardoso:2017kgn,Day:2019bbh}, however, in our work we mainly focus on the superradiance around a black hole.}. This is a phenomenon in which bosonic clouds are formed around a spinning black hole~\cite{Bekenstein:1973mi, Bekenstein:1998nt}. These clouds would then grow exponentially at the expense of the angular momentum and mass of the black hole~\cite{Detweiler:1980uk, Cardoso:2005vk, Dolan:2007mj}. In this work we consider these bosons to be massive dark photons forming clouds around the spinning primordial black holes (PBH). In addition to that we consider a special case of black hole superradiance, i.e., the quenched superradiance, where for a significant part of this dynamics, energy is drawn from the bosonic cloud due to the interactions of these bosons with other fermions. In this scenario, the creation of bosonic cloud continues for much longer due to the quenching as opposed to the vanilla black hole superradiance where due to the runaway growth of the cloud the angular momentum of the black hole reduces very fast making the whole process unsustainable. The main motivation is that the effective BSM interactions which can generate neutrino moments, can also create a stream of neutrinos from these dark photon clouds in the quenched superradiance paradigm~\cite{Chen:2023vkq, Banerjee:2024nga}. We lay down a foundation of a complementary probe of neutrino dark magnetic moments through the neutrino flux generated indirectly from PBH superradiance. 
This article is organized as follows, in Sec.~\ref{sec:moments} we briefly discuss how neutrino moments can be generated from their effective interactions with the dark photon followed by Sec.~\ref{sec:generalprod} where we discuss the production mechanism of the neutrinos flux from PBH superradiance.
In Sec.~\ref{sec:results} we present our results and finally we summarise and conclude in Sec.~\ref{sec:sumnconcl}.

\section{Moments from a dark photon}
\label{sec:moments}
\begin{figure}[h]
  \centering
  \includegraphics[width=2.5cm]{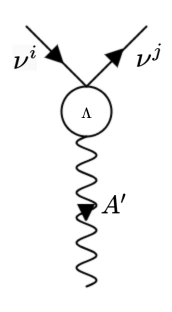}
  \caption{Neutrino dark photon one loop interaction.}
    \label{fig:feynman diagram}
\end{figure}

Electromagnetic interactions of neutrinos provide a unique avenue for the search of new physics. These spin flipping interactions of neutrinos were initially suggested as a solution  to the solar deficit problem as they could transform left handed active neutrinos to right handed sterile ones, thus accounting for the deficit \cite{Cisneros:1970nq,Lim:1987tk,Pulido:1990ba,Pulido:1991fb,Akhmedov:2000fj}. Even though with the discovery of neutrino oscillation this became a sub-leading effect, research in this field is active and is in fact very much essential as they could provide answers to many interesting questions in the field of neutrino physics such as the Dirac/Majorana confusion. While Dirac neutrinos possess both diagonal and transition moments Majorana neutrinos being their own antiparticles possess only transition moments since their electromagnetic interaction matrix is antisymmetric \cite{Schechter:1981hw,Pal:1981rm,Nieves:1981zt,Shrock:1982sc,Kayser:1984ge,Bilenky:1987ty, Nowakowski:2004cv, Wong:2005pa, Studenikin:2008bd,Giunti:2008ve,Broggini:2012df, Giunti:2014ixa,Akhmedov:2014kxa}. Thus an experiment measuring a non-zero diagonal electromagnetic moment of neutrino would confirm the nature of neutrino as Dirac.
Effective field theories with higher dimensional operators provide a framework to describe neutrino-photon couplings. In this work we consider the idea that the neutrinos obtain the moments through an effective interaction with a dark photon \cite{Herrera:2024iye}. The dark photon $A^{\prime}$ is the gauge field of the BSM $U(1)$ gauge symmetry indicating a dark sector and $F^{\prime\mu\nu}$ is the field strength corresponding to $A^{\prime}$. In our model-independent study, we consider an effective one loop interaction of neutrinos with massive dark photons which, as we discuss in the subsequent sections, take part in the black hole superradiance. The Feynman diagram for the same is shown in Fig.\ref{fig:feynman diagram}. 
The standard electromagnetic interaction of neutrinos can be described by the effective Hamiltonian,
\begin{equation}
\mathcal{H}(x) = j_{\mu}^{(\nu)}(x)\, A^{\mu}(x),
\end{equation}
where $A^{\mu}(x)$ is the electromagnetic field and the effective neutrino current $j_{\mu}^{(\nu)}(x)$ is given by,
\begin{equation}
j_{\mu}^{(\nu)}(x) = \sum_{k,j} \bar{\nu}_k(x)\,\hat{\Lambda}^{\mu}_{kj}\,\nu_j(x).
\end{equation}
Here $\hat{\Lambda}^{\mu}_{kj}$ is the effective electromagnetic operator and $k,j=1,2,3, \ldots$, is the sum over the indices of the massive neutrinos. In the SM scenario these indices go up to $k,j = 3$ but this could be extended with sterile neutrinos in extensions of SM. 
The vertex function $\Lambda$, shown in Fig. \ref{fig:feynman diagram}, arises when the effective neutrino current is sandwiched between the initial and final neutrino states. In general, it is a $4\times 4$ matrix in the spinor space and an apparent choice of its form would be to express it as a sum of product of Dirac $\gamma$ matrices. Thus the vertex function encompassing the electromagnetic properties of neutrinos takes the form~\cite{Nieves:1981zt,Kayser:1982br,Kayser:1984ge,Nowakowski:2004cv,Broggini:2012df,Chu:2018qrm,Giunti:2019aiy,Ibarra:2022nzm},
\begin{equation} 
\label{lambda}
    \Lambda^{\mu}_{A^{\prime}}=\left(\gamma^{\mu}-\frac{q^{\mu}\slashed{q}}{q^2}\right) [f_{Q}(q^2)+f_{A}(q^{2})q^{2}\gamma_{5}]-i\sigma^{\mu\nu}q_{\nu}[f_{M}(q^{2})+if_{E}(q^{2})\gamma_{5}].
\end{equation}
The subscript $A^{\prime}$ indicates that $f_{Q}$, $f_{A}$, $f_{M}$ and $f_{E}$ are the dark counterparts of the standard electromagnetic form factors. In the zero momentum limit these form factors become the dark charge, dark anapole, dark magnetic, and dark electric moment respectively (for e.g. $f_M(0)=\mu$ is the dark magnetic moment). Moreover, these can be translated to the standard electromagnetic moments through the kinetic mixing parameter.
The effective Lagrangian for the cases where the couplings are the dark moments can be given by~\cite{Herrera:2024iye},
\begin{equation}\label{laga}
    \mathcal{L}_{a^{\prime}}=\frac{a^{\prime}_{ij}}{2}\bar{\nu_{j}}\gamma_{\mu}\gamma_{5}\nu_{i}\partial_{\nu}F^{\prime\mu\nu}
\end{equation}
and
\begin{equation}\label{lagmu}
    \mathcal{L}_{\mu^{\prime}}=\frac{\mu^{\prime}_{ij}}{2}\bar{\nu_{j}}\sigma_{\mu\nu}\nu_{i}F^{\prime\mu\nu}.
\end{equation}
Here $a^{\prime}_{ij}$ and $\mu^{\prime}_{ij}$
are the dark anapole and magnetic moments, respectively and $\nu_{j}$, $\nu_{i}$ are the active neutrino states. In this work we take into consideration only the neutrino magnetic and anapole moment. This is because for ultrarelativistic neutrinos the electric dipole and magnetic moment are experimentally indistinguishable since they enter the observables as an equal combination of the two and also the anapole and charge radius contributions are phenomenologically the same \cite{Giunti:2014ixa}.
In the following we discuss the quenched superradiance of black hole arising from the effective couplings shown in Eqs.~\eqref{laga},~\eqref{lagmu}.
\section{Quenched superradiance}
\label{sec:generalprod}
The phenomenon of superradiance has emerged as an unique astrophysical laboratory providing stringent constraints on ultra light bosons and other extensions of the Standard Model. More recently the phenomenon of quenched superradiance has been explored where the process is prolonged due to the interaction of the boson cloud with fermions. In this section we discuss the quenching of the superradiant instability of the dark photon cloud through the interaction with neutrinos.
The dark photon generated via superradiance, is described by the following field~\cite{Detweiler:1980uk,Brito:2015oca,Baryakhtar:2017ngi,Siemonsen:2022ivj},
\begin{equation}\label{afield}
    A^{\prime \, \mu}=\Psi_{0}(t)e^{-\alpha_{g}^{2}r/r_{g}}(\alpha_{g}\sin\theta\sin(m_{A^{\prime}} t - \phi), \cos(m_{A^{\prime}} t), \sin(m_{A^{\prime}} t), 0),
\end{equation}
where $m_{A^{\prime}}$ is the mass of the dark photon, $r_{g}$ is the gravitational radius and $\Psi_{0}$ is the peak field value. The quantity $\alpha_{g}=\frac{G M_{\text{BH}}m_{A^{\prime}}}{\hbar c}$ is the gravitational fine structure constant where $G$ is Newton's gravitational constant, $M_\text{{BH}}$ the mass of the black hole and $m_{A^{\prime}}$ the mass of the vector boson relevant to the study.
It is important to look at the black hole spin for which superradiance can sustain. The growth stagnates when the mass of the cloud $(M_\text{{c}})$ is 10\% the mass of the BH $(M_\text{{BH}})$~\cite{Brito:2014wla}.  At this stage, the field value reaches a maximum,  close to the Planck scale~\cite{Chen:2022kzv}, and the spin of the BH reduces to a value such that it can no longer contribute to superradiance. The condition is given as,
\begin{equation}\label{spin}
\alpha_{g}<\frac{a_{*}}{2(1+\sqrt{1-a_{*}^{2}})},
\end{equation}
where $a_*$ is the black holes spin parameter that can be expressed as $a_* = J_{\rm BH}/GM_{\rm BH}^2$, where $J_{\rm BH}$ is the total angular momentum of the black hole.  

Superradiance stops when the spin of the black hole will not anymore be viable to maintain the continuous extraction of energy and angular momentum for the growth of the bosonic cloud. However, as mentioned before the longer sustenance of the phenomenon is possible by quenching the superradiance process via the interaction of the bosons with other standard model or beyond standard model particles. 
As introduced in Ref.~\cite{Chen:2023vkq} and Ref.~\cite{Banerjee:2024nga}, the cloud around the black hole can gain or lose mass through three main mechanisms which are, energy extraction of cloud from black hole by superradiance, energy loss of the cloud due to interaction with fermions and finally energy loss due to gravitational waves. As a result, the evolution of the bosonic cloud mass can be expressed as,
\begin{equation}\label{mastereqn}
\frac{dM_\text{c}}{dt} = \Gamma_{\text{SR}}M_{\text{c}} - 2 E_{f} \int \Gamma_{s/V} d^{3}x - \frac{dE_{\text{GW}}}{dt}.
\end{equation}
In the RHS of the above equation, the first, second, and third terms represent the superradiant growth of the cloud, the rate of energy loss due to production of fermions, and the energy radiated in the form of gravitational waves, respectively. It is to be noted that while superradiance continues, the third term is minuscule compared to the other two and therefore is ignored. As explained in subsequent parts of this article, the field value of the bosnic cloud needs to increase beyond a certain threshold for the fermionic production to begin. Therefore, in the first phase of superradiance, only the first term is active, i.e., the cloud grows without any significant loss of energy; this phase is termed as the \textit{growth phase}.
Then comes the \textit{balanced phase} when the fermionic production starts and balances the growth rate which leads to interesting physical phenomena like the production of diffferent SM and BSM particles. 
To get the duration of this phase, i.e., $\tau_{\mathrm{balanced}}$ we solve the black hole spin evolution equation $\frac{d a_*}{dt}=
-\frac{m_{\phi/A^{\prime}}\,\Gamma_{\mathrm {SR}}\,M_c}{\alpha_g^2}$ and find the duration till the condition in Eq. (\ref{spin}) is obeyed. Next, the gravitational atom enters the \textit{depletion phase} where superradiance has stopped, yet due to the existence of the high field value of the boson, fermionic production still continues for a while. Finally, once both the growth of the cloud and the production of fermions have stopped, there would be energy loss  only in the form of gravitational waves.
Among the three phases mentioned above, the balanced phase lasts the longest and the fermionic production is the most efficient in this phase. Therefore, we consider the neutrino production during the balanced phase.
In the following we discuss neutrino production from a single PBH as well as a neutrino background arising from a PBH population through the quenched superradiance mechanism.
\subsection{Neutrino production from point source} 
\label{subsec:production}
As mentioned in the previous section, during the balance phase of superradiance the bosons interact with fermions so that the rate of superradiance and rate of fermion production are balanced. 
These interactions ultimately lead to boosted fermion flux which could, in principle, be observed on earth. The fermion production from the dark photon happens through Schwinger pair production \cite{Schwinger:1951nm} and the rate is given as,
\begin{equation}\label{schwprodrate}
    \Gamma_{f}=\frac{g_{f}^{2}E_{A^{\prime 2}}}{48 \pi}.
\end{equation}
The production of these neutrinos requires the following condition to be satisfied,
\begin{equation}
    g_{f}E_{A^{\prime}}\gg m_{f}^{2},
\label{eq:prodcond}
\end{equation}
where $g_{f}$ is the coupling between the fermion and the dark photon, $E_{A^{\prime}}$ is the strength of the dark electric field and $m_{f}$ is the mass of the fermion. This translates into a more simplified form when we substitute $E_{A^{\prime}}\approx m_{A^{\prime}}\lvert \Vec{A^{\prime}}\rvert$ and Eq.~\eqref{afield} to Eq.~\eqref{eq:prodcond}. Thus the production criteria becomes,
\begin{equation}\label{constraintprod}
    \sqrt{g_{f}\Psi_{0}m_{A^{\prime}}}\gg m_{f}.
\end{equation}
In the balance phase of superradiance the field value $\Psi_0$, takes a critical value for which the superrradiant growth and energy loss are balanced. The expression for this critical value $\Psi^{\text{c}}_{0}$ is, ~\cite{Chen:2023vkq}, 
\begin{equation}\label{psicrit}
\Psi^{\text{c}}_{0}\approx 5.7 \times 10^{23}\left(\frac{1}{N_{f}}\right)\left(\frac{m_{A^{\prime}} \text{/eV}}{10^{-12}}\right)\left(\frac{\alpha_{g}}{0.3}\right)\left(\frac{10^{-12}}{g_{f}}\right)^{3}\left(\frac{\alpha_{J}}{0.9}\right).
\end{equation}
Here $N_{f}$ is the number of fermion flavor species that take part in the pair production.

The fermions produced are then accelerated by the `electric' field arising from the vector boson to energies proportional to the coupling $g_{f}$. The flux of these boosted fermions could be observed on earth through various detectors and telescopes. The energy and the differential flux of the fermions are given by~\citep{Chen:2023vkq},
\begin{equation}\label{genenergy}
E_{f}=0.35g_{f}\Psi_{0}
\end{equation}
and
\begin{align}\label{gendifflux}
    \frac{d\Phi_{f}}{dE} & = 1.3\times 10^{-6}\, \left(\frac{\Psi_{0}}{5.7\times 10^{14}\, \text{GeV}}\right)\left(\frac{N_{f}}{1}\right)\left(\frac{10^{-12}\,\text{eV}}{m_{A^{\prime}}}\right)\left(\frac{g_{f}}{10^{-12}}\right)\cr
   & \times \left(\frac{0.3}{\alpha_{g}}\right)^{3}\left(\frac{5\,\text{kpc}}{d}\right)^{2}\, \text{cm$^{-2}$ s$^{-1}$eV$^{-1}$},
\end{align}
respectively. Here $d$ is the distance of the BH from the observer in kpc.
Since in this work we consider neutrino production, it is essential to discuss the effective neutrino coupling involved in neutrino-dark photon interaction ($g_\nu \bar{\nu}\gamma_{\mu}A^{\prime \mu}\nu$). Taking into consideration the Lagrangians in Eqs.~\eqref{laga} and \eqref{lagmu}, and the form of the gauge field in Eq.~\eqref{afield} the coupling would take the form,
\begin{equation}\label{coupling}
    g_{\nu} = m_{A^\prime}^{2}\, a^\prime + m_{A^\prime}\, \mu^\prime .
\end{equation}
Here $a^\prime$ and $\mu^\prime$ are the dark anapole and magnetic moments when only one active neutrino state is considered. In our study we consider it to be the electron neutrino, $\nu_{e}$. These dark moments translate to standard electromagnetic moments through the kinetic mixing parameter i.e., $a^\prime = \frac{a}{\epsilon}$ and $\mu^\prime = \frac{\mu}{\epsilon}$ where $a$ and $\mu$ are the standard electromagnetic anaploe and dipole moments of the neutrinos. Various theoretical, experimental, and observational bounds on the mixing parameter can be found in Ref.~\cite{Fabbrichesi:2020wbt}.

The product of the masses of the dark photon and the black hole obey the condition\footnote{This is derived from considering $\alpha_g\sim\mathcal{O}(0.3)$ which we use throughout this article.};
\begin{equation}\label{constraintBH}
   m_{A^{\prime}}M_{\text{BH}}\approx 10^{-20},
\end{equation}
where ${M}_{\text{BH}}$ is in the unit of solar mass $M_{\odot}$, and  $m_{A^{\prime}}$ is in GeV.
The constraints on the mass range of the dark photon taken here comes from two considerations.
\begin{enumerate}
    \item It has been shown that black holes with near extremal rotation below a certain mass, i.e., $2.5\times10^{-19}\, M_{\odot}$~\cite{Laha:2020vhg}, has already evaporated. From this we get the upper bound on $m_{A^{\prime}}$ to be $20\,\text{MeV}$ by taking ${M}_{\text{BH}}=2.5\times10^{-19}{M}_{\odot}$.
    \item The lower bound on the dark photon mass is derived from the Schwinger pair production condition given in Eq. (\ref{constraintprod}). Here it is to be noted that we will have a particular lower bound value of the dark photon mass for a fixed $g_{\nu}$. From Eq. (\ref{constraintprod}) and the expression for critical field value given in Eq. (\ref{psicrit}) we arrive at the following coupling dependent lower mass bound for the dark photon,
\begin{equation}
m_{A^{\prime}}\gg 265g_{\nu} \, \text{eV}.
\end{equation}
The values for the standard anapole and magnetic moment are taken from the bounds set by the current XENONnT electron recoil data~\cite{XENON:2022ltv}. The limit set on the anapole moment is $a\leq7.5\times10^{-5} \,\text{GeV}^{-2}$ and on the magnetic moment is $\mu\leq5.66\times10^{-9}\,\text{GeV}^{-1}$.
\end{enumerate}

\begin{figure}[t]
\centering
\includegraphics[scale=0.59]{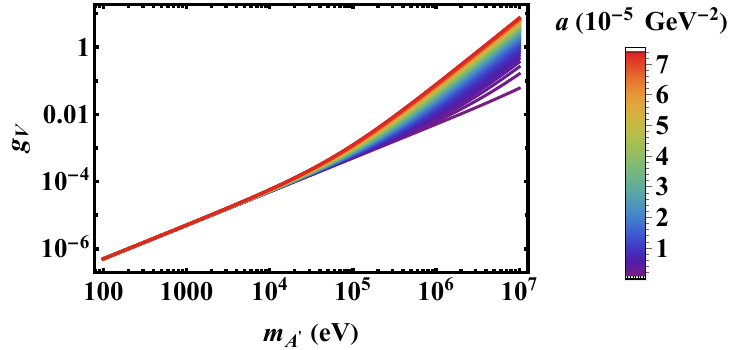}
\includegraphics[scale=0.59]{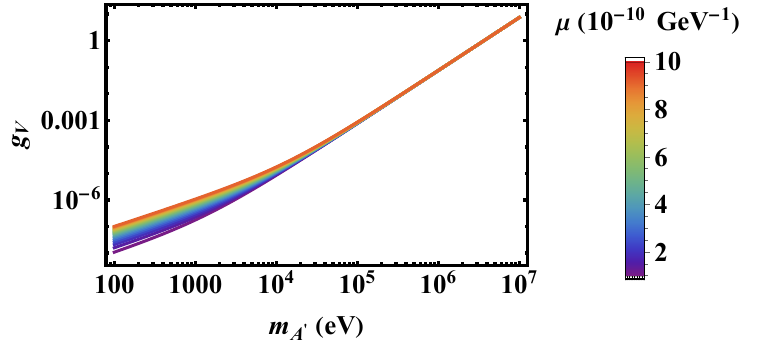}
\caption{Left: Dependence of $g_V$ on $m_A$ for different $a$ when $\mu=5\times 10^{-9}\mathrm{~GeV^{-1}}$. Right: Dependence of $g_V$ on $m_A$ for different $\mu$ when $a=7.45\times 10^{-5} \mathrm{~GeV^{-2}}$.} 
\label{gvvsma}
\end{figure}

According to the conditions and the bounds mentioned above, the effective coupling between the neutrinos and the dark photon can have values within a certain range which we show in Fig.~\ref{gvvsma}.
It can bee seen from the figure that at the lower end of the dark photon mass the value of the magnetic moment dominates the effective coupling and at the higher end of the dark photon mass the anapole moment dominates the effective coupling. 

\subsection{Diffuse quenched superradiance neutrino background} \label{dqsnb}
Hitherto we discussed the production of neutrinos from an isolated point PBH source. As an extension to what has already been discussed, here we take up the case of the production of a neutrino background from a primordial black hole population.  
Primordial black holes unlike the astrophysical black holes trace back their origin to very high red-shifts. These PBHs with sufficiently high initial spin can produce a diffuse neutrino background through the mechanism of quenched superradiance, which we call the diffuse quenched superradiance neutrino background (DQS$\nu$B). The observables concerned with this background created in the early Universe would be significantly red-shifted and the formalism for this and concepts discussed in this section can be found in Ref.~\cite{Banerjee:2025ddr}. 
The observed diffuse neutrino flux from quenched superradiance can be expressed as, 
\begin{equation}\label{finalphi}
\frac{d\Phi_{\text{obs}}}{dE} =
\begin{cases}
    cn_{(\text{PBH,0})}\frac{1+z_{*}}{H(z_{*})}\frac{R_{\text{tot}}}{E_{\text{obs}}} & z_{\text{lower}} \leq z_{*} \leq z_{\text{upper}}, \\
    0 & \text{everywhere else}.
\end{cases}
\end{equation}
Here, $H(z_{*})$ is the Hubble parameter at redshift $z_*$, $R_{\text{tot}}\equiv E_{\text{so}}\frac{d \phi}{dE_{s}}(4\pi d^{2})$ is the neutrino flux from an individual source in its source frame where $E_{\text{so}}=E_f$ is the fixed energy of the neutrinos at the source frame and $E_{\text{obs}}$ is the energy of the neutrinos in the observer frame. The present day primordial black hole number density is, $n_{(\mathrm{PBH},0)}\equiv \frac{f_{\text{PBH}}\rho_{\text{DM,0}}}{\text{M}_\text{PBH}}$, where $f_{\text{PBH}}$ is the primordial black hole abundance and it indicates the fraction of dark matter constituted by PBH; $\rho_{\text{DM,0}}$ is the dark matter density in the present universe. The limits $z_{\rm upper}$ and $z_{\rm lower}$ correspond, respectively, to the onset of the balanced phase and the depletion phase of quenched superradiance. For simplicity, we assume monochromatic mass spectrum of PBH, i.e., all PBHs in the population to have originated with the same mass. Consequently all the neutrinos are created with the same source frame energy $E_{\rm so}$. However, due to the time difference between the commencement and the conclusion of the balanced phase, the energy of the neutrinos produced at different times get redshifted differently, leading to a broad energy range. The energy interval of this diffuse flux is in between $\frac{E_{f}}{1+z_{\text{upper}}}$ and $\frac{E_{f}}{1+z_{\text{lower}}}$.
In Ref.~\cite{Banerjee:2025ddr} lower bounds on the coupling between neutrinos and the bosons involved in the quenched superradiance mechanism were obtained. This was done using the constraints on $\Delta N_{\mathrm{eff}}$, flux bounds from Super-Kamiokande while assuming the maximum allowed values of $f_{\rm PBH}$ for different masses of the PBHs. In this work, however, we employ the existing bounds on the electromagnetic moments of neutrinos to obtain the bound on $f_{\rm PBH}$ from $\Delta N_{\rm eff}$. 
\subsubsection*{Bounds on the PBH abundance}\label{fpbhbounds}
In standard cosmology $N_{\mathrm{eff}}$ indicates the number of relativistic degrees of freedom contributing to the radiation energy density in the early Universe. The difference between the standard model prediction and Planck measurement of $N_{\mathrm{eff}}$ corresponds to the excess radiation density which could arise due to any process motivated by new physics, such as the neutrino production mentioned in this work. Ref.~\cite{Lin:2026yov} discusses the galactic and extragalactic contributions to the neutrino flux while deriving bounds on PBH dark matter. In this work $f_{\text{PBH}}$ bounds are derived solely from $\Delta N_{\mathrm{eff}}$ where only the pre-recombination extragalactic flux is relevant since the galactic component does not contribute to the radiation energy density at recombination. The expression for the difference in $N_{\mathrm{eff}}$ at the time of recombination is given by,
\begin{equation}
\label{deltaNgen}
\Delta N_{\mathrm{eff}}=\frac{8}{7}\left(\frac{11}{4}\right)^{\frac{4}{3}} \frac{\rho_{\nu}^{\mathrm{extra,now}}}{\rho_{\gamma}^{\mathrm{now}}}.
\end{equation} 
Here $\rho_{\gamma}^{\mathrm{now}}$ is the present photon energy density. The present excess radiation density in the form of neutrinos is $\rho_{\nu}^{\mathrm{extra,now}}=\frac{4 \pi}{c} \int_{0}^{\infty} E\frac{d\Phi_{\mathrm{pre}}}{dE}\,dE$.
It is to be noted that neutrinos produced via quenched superradiance after the recombination epoch does not contribute to the $\Delta N_{\rm eff}$, hence we do not take that into account. Thus by substituting $\rho_{\nu}^{\mathrm{extra,now}}$ back into Eq. \eqref{deltaNgen} we have the expression,
\begin{equation}
\label{deltaN}
\Delta N_{\mathrm{eff}}= 7.083 \times 10^{-9} \int_{0}^{\infty} E\frac{d\Phi_{\mathrm{pre}}}{dE}\,dE.
\end{equation}
Since the diffuse neutrino flux depends explicitly on the primordial black hole abundance, $\Delta N_{\mathrm{eff}}$ is also a function of $f_{\mathrm{PBH}}$. We therefore derive bounds on $f_{\mathrm{PBH}}$ by requiring that $\Delta N_{\mathrm{eff}}\leq 0.3$ which is the difference in the value of $N_{\mathrm{eff}}$ measured by the Planck (2018) data \cite{Planck:2018vyg} and the standard model predicted value \cite{Mangano:2005cc,Cielo:2023bqp} .

\section{Results}
\label{sec:results}
In the initial part of this section, we discuss the possible neutrino flux arising from point PBH sources for various benchmark cases. Furthermore, in the second part of this section we show the results for the idea discussed in section \ref{dqsnb}. Here, instead of focusing on neutrino flux from a point source black hole we take a diffuse flux from a population of primordial black holes. The resulting flux is constrained using $\Delta N_{\mathrm{eff}}$ bounds which along with the existing upper limits on the neutrino electromagnetic moments, enables us to derive corresponding upper bounds on the primordial black hole abundance, $f_{\mathrm{PBH}}$. It is to be noted that in all subsequent discussions in this section, we consider the initial value of the PBH spin parameter to be $a_* = 0.9$. Such an extremal spin can naturally arise if these black holes are formed during an early matter-dominated epoch from the collapse of asymmetric overdensities. Violent bubble collisions during first-order cosmological phase transitions can also generate these rapidly rotating primordial black holes.
%

\subsection{Isolated black hole source}
In the above discussion we have explained how to obtain the neutrino energy and flux along with the duration of the neutrino production event through quenched superradiance. In our results we aim to focus on benchmark parameters that lead to one or more detectable ultra high energy neutrinos on earth. Therefore, the important quantity here is the number of events which can be expressed as,
\begin{align}
N_{\mathrm{events}} = E_{f}\dfrac{d\Phi_{f}}{dE}\tau_{\mathrm{balanced}}A_{\mathrm{det}},
\end{align}
where $A_{\mathrm{det}}$ is the effective area of the detector. In this study we consider the effective area to be $A_{\mathrm{det}}=1\mathrm{~km^2}$ to calculate the number of events. The choice is motivated by the characteristic detector area of large scale neutrino telescopes such as IceCube and KM3NeT (ARCA) which are discussed in later parts of this section. 
We show benchmark cases for which we get $N_{\mathrm{events}} \geq 1$ where all the constraints on the various parameters are maintained in Table~\ref{regIbp}. 
\begin{table}[H]
\centering
\begin{tabular}{|c|c|c|c|c|c|c|}
\hline
 BP & $m_{A^{\prime}}$ (GeV) & $M_{\mathrm{PBH}}~(M_{\odot})$ & $a~(\mathrm{GeV}^{-2})$ & $\mu~(\mathrm{GeV}^{-1})$ & $d~(\mathrm{kpc})$ & $N_{\mathrm{events}}$ \\ \hline \hline
1 & $2.5\times 10^{-7}$ & $4\times 10^{-14}$ & $3.75\times 10^{-6}$ & $2.5\times 10^{-10}$ & $10^4$ & $24$ \\ \hline
2 & $1.5\times 10^{-6}$ & $6.67\times 10^{-15}$ & $2.48\times 10^{-6}$ & $1.25\times10^{-10}$ & $5\times 10^3$ & $25$ \\ \hline
3 & $1.53\times 10^{-5}$ & $6.53\times 10^{-16}$ & $9.31\times 10^{-8}$ & $6.25\times 10^{-12}$ & $1050$ & $2.01$ \\ \hline
4 & $9\times 10^{-4}$ & $1.11\times 10^{-17}$ & $8.27\times 10^{-10}$ & $5.55 \times10^{-14}$ & $150$ & $1$ \\ \hline
5 & $5\times 10^{-3}$ & $2\times 10^{-18}$ & $2.32\times 10^{-11}$ & $3 \times10^{-14}$ & $24$ & $1$ \\ \hline
6 & $1.1\times 10^{-2}$ & $9.09\times 10^{-19}$ & $9.43\times 10^{-12}$ & $8.33 \times10^{-16}$ & $20$ & $1$\\ \hline
7 & $2\times 10^{-2}$ & $5\times 10^{-19}$ & $8.27\times 10^{-12}$ & $7.6 \times10^{-16}$ & $15$ & $5$\\ \hline
\end{tabular}
\caption{Details of the benchmark parameters. The Milky Way galaxy has a diameter of around 30 kpc and hence BPs 5, 6, and 7 represent possible PBHs residing within our galaxy.}
\label{regIbp}
\end{table}

\begin{figure}[t]
\centering
\includegraphics[scale=0.6]{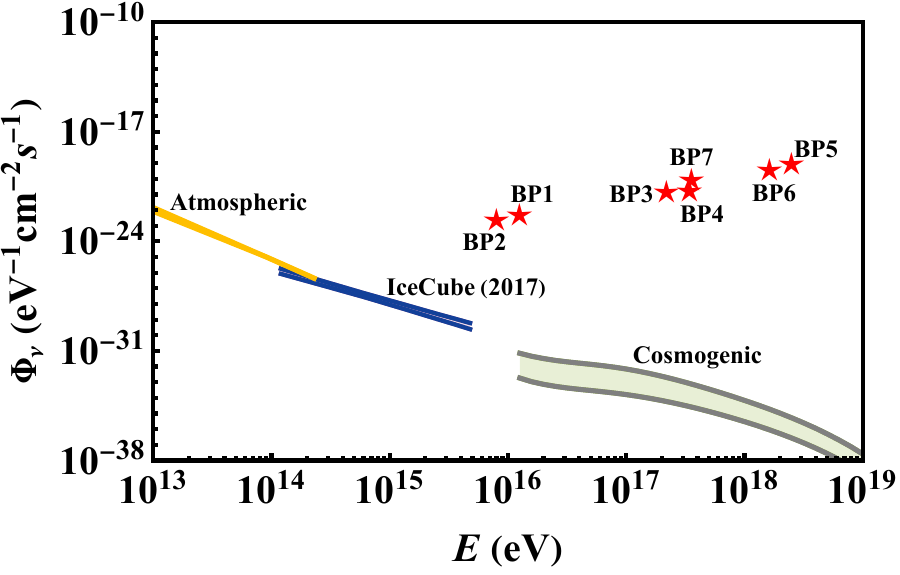}
\caption{Neutrino flux vs. energy for the benchmark points given in Table \ref{regIbp}. The atmospheric, IceCube and cosmogenic neutrino flux shown in the plot is taken from \cite{Vitagliano:2019yzm}.}
\label{bpflux}
\end{figure}

In Fig. \ref{bpflux}, we show the outcome of the BPs mentioned above in the neutrino energy and flux plane where we also mention the number of events detectable on earth for each case. Apart from the fluxes shown in Fig. \ref{bpflux}, possible probes for the ultra high energy neutrino fluxes arising from the bench mark parameters under consideration include, IceCube-Gen2 \cite{IceCube-Gen2:2020qha}, RNO-G \cite{RNO-G:2020rmc} and GRAND200k \cite{GRAND:2018iaj}. BP 1 and 2 correspond to PBHs located at distances $10^4$ kpc and $5 \times 10^3$ kpc respectively, whose neutrino fluxes fall within the sensitivity range reported by IceCube \cite{IceCube:2017zho}.
We now discuss BPs 3, 4 and 7 in detail since these BPs point to neutrino energies and number of events similar to that of the KM3NeT highest neutrino energy event, namely KM3-230213A \cite{KM3NeT:2025npi}. The KM3NeT collaboration announced the detection of the most energetic neutrino event at the 21-line configuration of its ARCA detector recently. This single event estimates the neutrino energy to be around $220$ PeV, at least 10 orders of magnitude higher than that measured by IceCube \cite{IceCube:2013cdw,IceCube:2021rpz,IceCubeCollaborationSS:2025jbi}.
The specific BPs mentioned above have energies in the range $221-360$ PeV and the number of events varies from 1-5. Hence, in the light of the KM3NeT event, production of neutrinos by the quenched superradiance of primordial black holes will be an interesting phenomenon to explore. We leave the interesting possibility of an explanation for the KM3NeT event by employing a detailed statistical analysis in the current framework for future study.
%
%
%
%
\subsection{$f_{\text{PBH}}$ bounds from DQS$\nu$B}
In section \ref{fpbhbounds}, we discussed the idea of constraining the abundance of primordial black holes $f_{\text{PBH}}$ through $\Delta N_{\mathrm{eff}}$ bounds. Ref.~\cite{Banerjee:2025ddr} discussed both the cosmological  as well as the Super-Kamiokande bounds while deriving the bounds on neutrino couplings. In this analysis we do not include the latter since the Super-Kamiokande bound requires the maximum observable neutrino energy to exceed the detector threshold of 17.3 MeV. For the coupling considered here, in the bulk of our parameter space, the upper limit of the observed neutrino energy lies well below this threshold.  

\begin{figure}[t]
\centering
\includegraphics[scale=0.6]{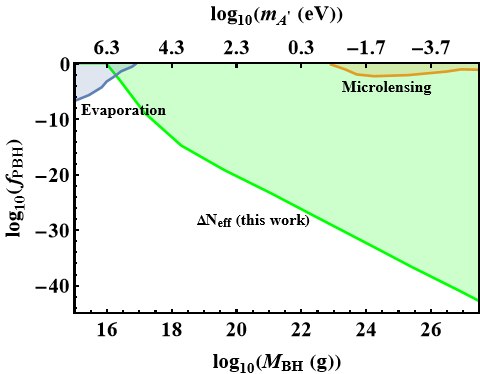}
\caption{Constraints on $f_{\text{PBH}}$ for varying mass of black holes. The constraints shown in this figure in the form of evaporation and microlensing bounds are taken from \cite{Carr:2020gox}.}
\label{fpbhmaxplot}
\end{figure}

\begin{figure}[H]
\centering
\includegraphics[scale=0.25]{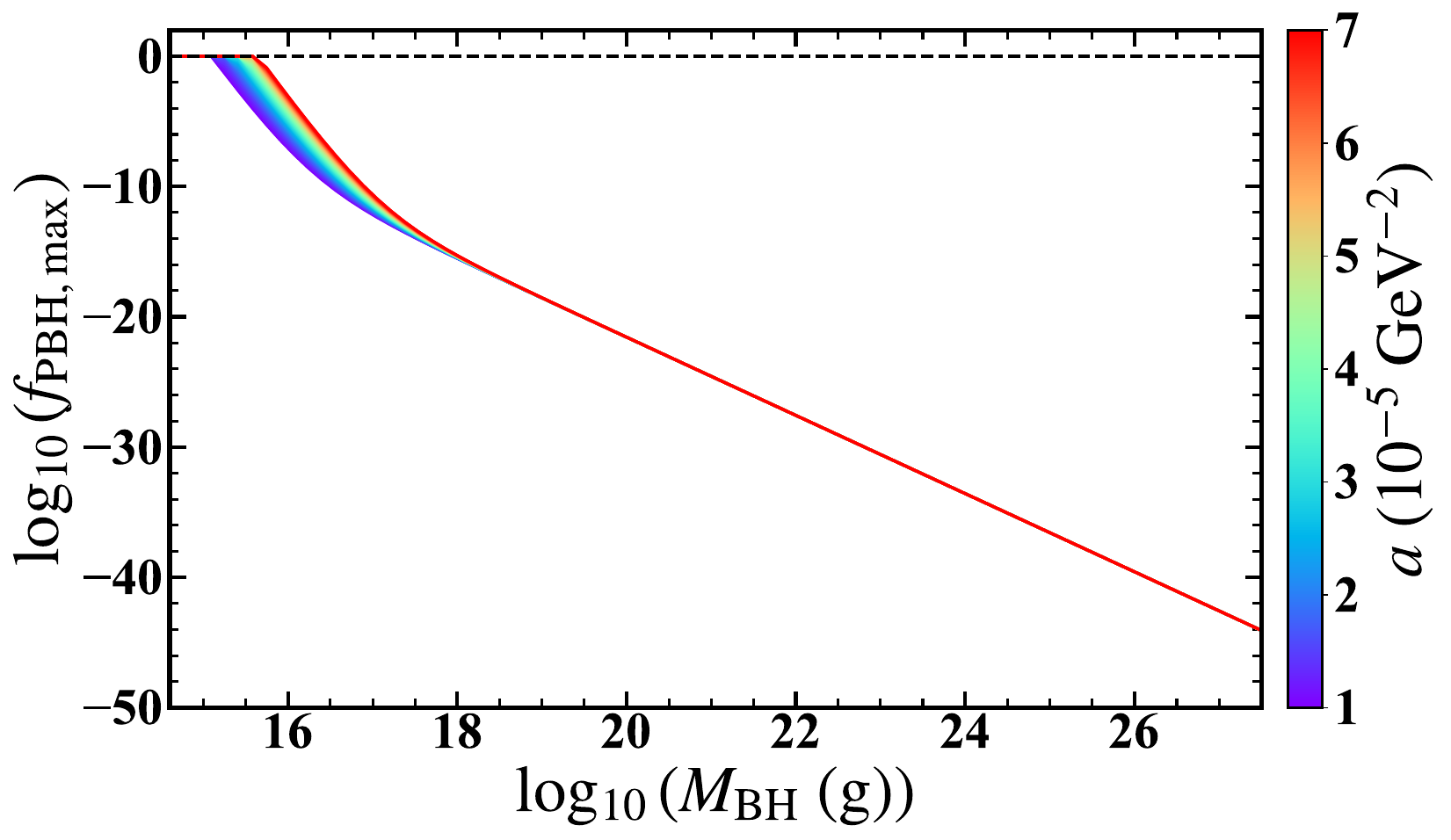}~~
\includegraphics[scale=0.25]{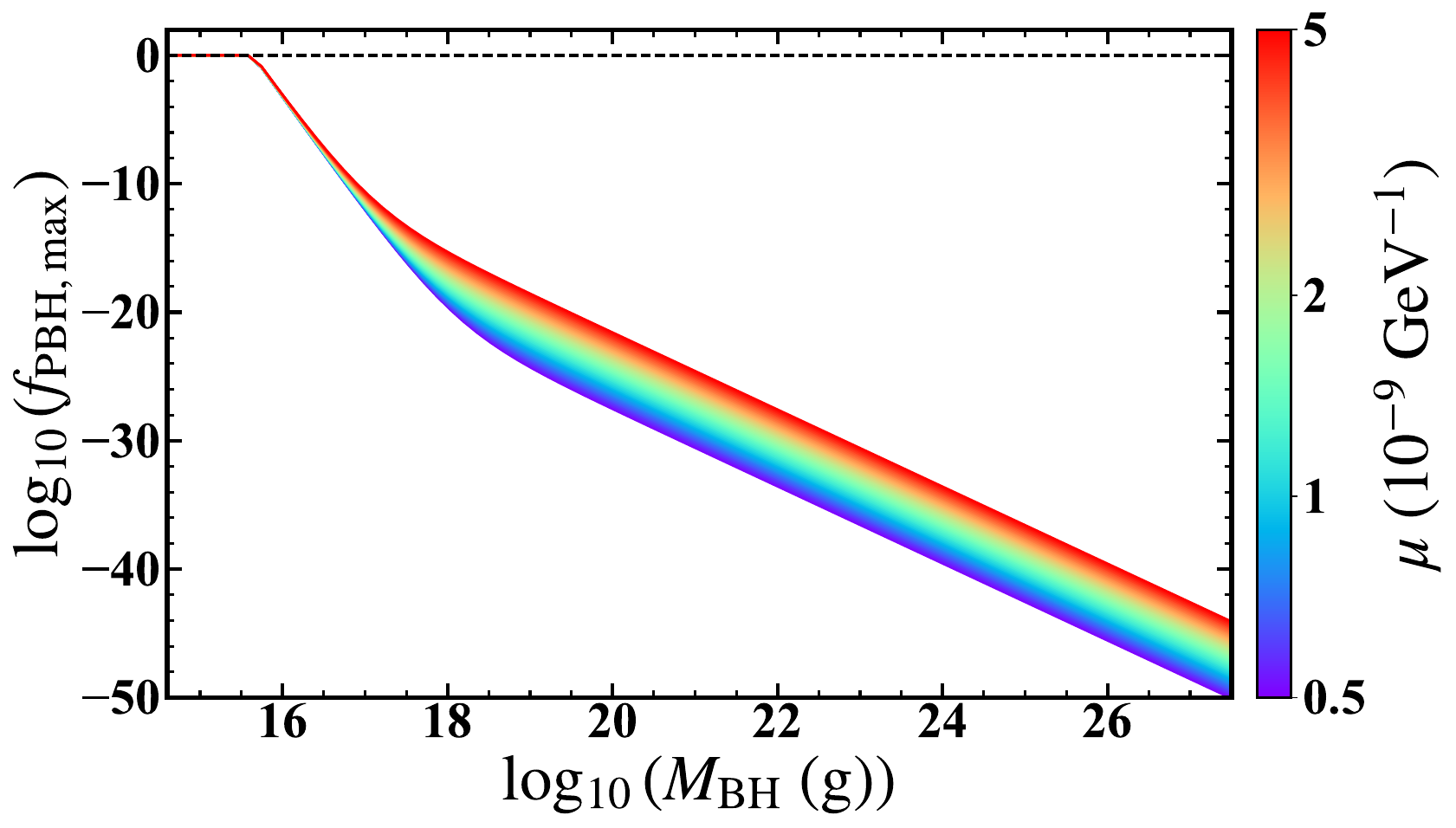}
\caption{Left: Dependence of $f_{\text{PBH}}$ on $m_A$ for different $a$ when $\mu=5\times 10^{-9}\mathrm{~GeV^{-1}}$. Right: Dependence of $f_{\text{PBH}}$ on $m_A$ for different $\mu$ when $a=7.45\times 10^{-5} \mathrm{~GeV^{-2}}$.}
\label{fpbhplots}
\end{figure}

In Fig. \ref{fpbhmaxplot} we show the maximum allowed values of $f_{\text{PBH}}$ such that for the maximum values of magnetic and anapole moments the $\Delta N_{\mathrm{eff}}$  bound is satisfied.
Now, we vary these moments individually and obtain the corresponding maximum allowed abundance for different values of the dark photon mass which ultimately correspond to different PBH mass by fixing the gravitational fine structure value to $\alpha_g=0.3$. This is shown in Fig. \ref{fpbhplots}.
We can see features similar to Fig. \ref{gvvsma}, since the abundance would have a dependency directly proportional to the effective coupling. The difference appearing here is because the $x$-axis here is mass of the black hole instead of mass of the dark photon. This is intuitive because, as the coupling decreases the critical field value and hence the flux increases. Thus, in order for the $\Delta N_{\mathrm{eff}}$ to be maintained the abundance of primordial black holes must decrease in order to reduce the flux.

\section{Summary and Conclusion}
\label{sec:sumnconcl}
In this article we have developed a framework for probing the dark electromagnetic moments of neutrinos through the phenomenon of quenched black hole superradiance. We extended the idea of superradiant instability of a massive dark vector cloud into an astrophysical probe of dark neutrino magnetic and anapole moments.
By constructing the loop level effective vertex connecting dark photon to the  active neutrinos through the dark magnetic and dark anapole moment operators, we obtained an effective coupling whose strength depends on the dark photon mass. 
We bound the mass of the dark photon from above to $20\mathrm{~MeV}$ and showed the explicit dependence of these couplings on the dark photon mass. We found the possible observable fluxes for various benchmark parameters, some of which resemble the KM3-230213A, the highest neutrino energy event. 
In one of our previous studies we analysed a diffuse neutrino background generated by quenched superradiance of primordial black holes, where we derived lower bounds on generic scalar and vector boson couplings to neutrinos through the requirement of efficient quenching. In the current work, rather than treating the boson–neutrino coupling as a free parameter, we embedded it in the neutrino electromagnetic moment framework. Using this prescription along with the $\Delta N_{\mathrm{eff}}$ constraints, we derive bounds on the cosmological abundance of primordial black holes hosting such superradiant clouds. Thus this work effectively demonstrates that bounds on the neutrino electromagnetic moments provide novel means of constraining the primordial black hole abundance. In summary, in this work we have introduced quenched superradiance as a novel probe of  electromagnetic moments of neutrinos and we also derived cosmological bounds on primordial black hole abundance.
As a future exploration, the robustness of the quenching derived bounds against a non-minimal dark sector construction can be tested, for instance scenarios in which the dark photon also couples to a dark fermionic species that can absorb energy from the cloud. 
The projected sensitivities of upcoming CMB experiments such as CMB-S4 and Simons Observatory are expected to tighten the $\Delta N_{\mathrm{eff}}$ bound by roughly an order of magnitude relative to the current Planck limits, offering an avenue for refining the primordial black hole abundance bound derived here. 
%

%

\acknowledgments{UKD acknowledges support from the Anusandhan National Research Foundation (ANRF), Government of India under Grant Reference No.~CRG/2023/003769. AJ thanks the DST-INSPIRE for support through the INSPIRE fellowship. AJ thanks Shamik Niyogi for useful discussions.}



\bibliographystyle{JHEP}
\bibliography{numomSRref1.bib}

\end{document}